\documentstyle[preprint,eqsecnum,aps]{revtex}

\newcommand{\comm}[1]{\left[#1\right]}
\newcommand{\boldnabla}{\mbox{\boldmath$\nabla$}}

\newcommand{\boldsigma}{\mbox{\boldmath$\sigma$}}
\newcommand{\boldtau}{\mbox{\boldmath$\tau$}}

\begin{document}
\draft

\title{Pedagogical Studies of Relativistic Hamiltonians}

\author{J. L. Forest and V. R. Pandharipande}
\address{Department of Physics, University of Illinois at Urbana-Champaign,
         1110 W. Green St., \\
Urbana, IL 61801}
\author{J. L. Friar}
\address{Theoretical Division, Los Alamos National Laboratory, MS B283,
         Los Alamos, NM 87545}

\date{September 21, 1994}
\maketitle
\begin{abstract}
Relativistic Hamiltonians are defined as the sum of relativistic
one-body kinetic energies, two- and many-body interactions and their boost
corrections.  We review the calculation of the boost correction of the
two-body interaction from commutation relations of the Poincar\'{e} group and
show that its important terms can be easily understood from classical
relativistic mechanics.  The boost corrections for scalar- and vector-%
meson-exchange interactions, obtained from relativistic field theory, are
shown to be in agreement with the results of the classical calculation. These
boost corrections are also shown to be necessary to reproduce the known
results of relativistic mean-field theories. We conclude with comments on the
relativistic boost operator for the wave function of a nucleus. Several of the
results presented in this pedagogical article are known. We hope that a better
understanding of relativistic Hamiltonians and their relation to relativistic
field theory is obtained by putting them together with a few new relations.
\end{abstract}
\pacs{\ \ \ PACS number: 21.30.+y}

\newpage
\section{INTRODUCTION}
The concept of an interparticle potential has proved to be extremely useful
in the study of non-relativistic many-particle systems at low energies.
The degrees of freedom associated with the fields coupled to the particles,
as well as the internal degrees of freedom, if any, of the particles are
eliminated with these potentials so that one can focus on the most important
degrees of freedom. ${\it Ab\ initio}$ calculations of the interparticle
potentials are non-trivial, particularly when the particles are composite like
nucleons or rare gas atoms. In practice the potentials are parameterized
within a suitable theoretical framework and fitted to observed data.
Non-relativistic Hamiltonians of the type:
\begin{equation}
   H_{NR}=\sum_i\frac{p_i^2}{2m_i}+\sum_{i<j}v_{ij}+\sum_{i<j<k}V_{ijk}+\
\cdots
\end{equation}
have been used in many contexts. In nuclear physics, for example, the ground
and low-energy nuclear states are described by eigenfunctions
$\Psi_{I}(x_1,x_2,\cdots,x_{A})$ of $H_{NR}$, where $x_i$ denotes
the position ${\bf r}_i$, spin $\boldsigma_i$ and isospin $\boldtau_i$ of
the $i$th nucleon.  Solving the many-body Schr\"{o}dinger's
equation:
\begin{equation}
   H_{NR}\:\Psi_{I}=\:E_{I}\Psi_{I}
\end{equation}
is a difficult problem; however, it can now be solved with variational
\cite{wiringa} and Green's function\cite{carlson1} Monte Carlo (VMC and GFMC)
methods for the ground and some low-energy excited states of up to six
nucleons.

Bakamjian and Thomas\cite{bakamjian} and Foldy\cite{foldy} showed many years
ago that the concept of potentials can also be useful in describing
many-body systems in a relativistically covariant fashion. The relativistic
Hamiltonian may be written as:
\begin{eqnarray}
   H_{R}=\sum_i\sqrt{m_i^2+p_i^2}\ +\
\sum_{i<j}\left[\tilde{v}_{ij}+{\delta}v_{ij}({\bf P}_{ij})\right]\ +\
\sum_{i<j<k}\left[\tilde{V}_{ijk}+{\delta}V_{ijk}({\bf P}_{ijk})\right]\ +\
\cdots,
\end{eqnarray}
where $\tilde v_{ij}$ are two-body potentials in the ``rest frame'' of
particles $i$ and $j$ (i.e. the frame in which their total momentum vanishes):
\begin{equation}
{\bf P}_{ij}={\bf p}_i+{\bf p}_j=0.
\end{equation}
Similarly $\tilde V_{ijk}$ is the three-body potential in the frame in which
\begin{equation}
{\bf P}_{ijk}={\bf p}_i+{\bf p}_j+{\bf p}_k=0.
\end{equation}
The $\delta v_{ij}({\bf P}_{ij})$ and $\delta V_{ijk}({\bf P}_{ijk})$ are
called ``boost interactions'' and depend upon the total momentum of the
interacting particles. Obviously,
\begin{equation}
\delta v_{ij}({\bf P}_{ij}=0)=\delta V_{ijk}({\bf P}_{ijk}=0)=0.
\end{equation}
Only the positive value of $\sqrt{m_i^2+p_i^2}$ is considered in $H_{R}$.

The interaction $\tilde v$ is determined by the fields and the internal
structure associated with the interacting particles, while $\delta v({\bf P})$
is related to $\tilde v$ by relativistic covariance. Krajcik and
Foldy\cite{krajcik} formally calculated $\delta v({\bf P})$ to all orders in
$P^2/4m^2$, though we will retain only the leading contribution of order
$P^2/4m^2$ in this study. An elegant equation for the minimal first order
$\delta v({\bf P})$ is found to be\cite{friar}:
\begin{equation}
\delta{v}({\bf P})=-\frac{P^2}{8m^2}\tilde{v}+\frac{1}{8m^2}\,\left[\ {\bf
P}\cdot{\bf r}{\bf P}\cdot\boldnabla, \tilde{v}\
\right]+\frac{1}{8m^2}\,\left[\ (\boldsigma_1-\boldsigma_2)\times{\bf
P}\cdot\boldnabla, \tilde{v}\ \right],
\end{equation}
where the subscripts $ij$ of $\tilde v$, ${\bf P}$, ${\bf r}$ and $\boldnabla$
have been suppressed for brevity. A brief explanation of this equation is
given in section II for completeness.

Present interest in the relativistic Hamiltonian (1.3) stems from the fact
that its ground states can be studied with the variational Monte Carlo method.
Initially only the $^3$H and $^4$He ground states were studied\cite{carlson2},
but the methods developed there can also be used to study heavier nuclei
like $^{16}$O with cluster expansions\cite{pieper}. Moreover, these methods
can also be used to calculate the ground states of $^2$H, $^3$H, $^3$He,
$^4$He, $^6$He and $^6$Li exactly, up to order $P^2/4m^2$, with the Green's
function Monte Carlo method\cite{pudliner}. The results obtained in ref. [7]
show that the average value of $P^2/4m^2$ is rather small in nuclei; therefore
it is certainly useful to have exact results to this order. It is very likely
that higher-order contributions and $\delta V({\bf P}_{ijk})$ contributions are
much smaller. For the sake of brevity we will not discuss
$\delta V({\bf P}_{ijk})$; a two-pion exchange term in it has been studied by
Coon and Friar\cite{coon}.

Before commencing our pedagogical discussion, it is necessary to categorize
the various effects that we will be treating. It is convenient to
separate relativistic effects in nuclear physics into three categories:
(A) on the interaction $\tilde{v}_{ij}$ of two nucleons in their center of
mass (CM) frame; (B) on the interaction $\tilde{v}_{ij}+\delta v({\bf P}_{ij})$
of nucleons $i$ and $j$ with a total momentum ${\bf P}_{ij}$ in the CM frame of
the whole nucleus; and (C) on the motion of the nucleus as a whole.

The first effect (A) depends essentially on the nature of the interaction;
for example, when $\tilde{v}_{ij}$ is mediated by mesons the relativistic
corrections to it depend upon the type, i.e. scalar, vector etc., of the
exchanged meson. Since all realistic models of $\tilde{v}_{ij}$ are
obtained by fitting experimental data, they contain relativistic effects
in some form. The key question here is how to choose the theoretical form
of $\tilde{v}_{ij}$, used to fit the data, such that they are correctly
represented.

There is no further model dependence in the second effect (B). The
$\delta v({\bf P}_{ij})$ depends only upon $\tilde{v}_{ij}$ and can be obtained
from it\cite{krajcik,friar}. Many aspects of the relation between
$\delta v({\bf P}_{ij})$ and $\tilde{v}_{ij}$ can be understood from
classical relativistic mechanics as discussed in ref. [7] and further
elaborated in section III. This allows terms in $\delta v({\bf P}_{ij})$ to be
classified as those coming from the relativistic kinematics, Lorentz
contraction, Thomas precession and quantum effects respectively.

The relativistic effect (C) on the motion of the nucleons as a whole is
important because in scattering experiments the struck nucleus recoils
and phenomena such as Lorentz contraction, Thomas precession, and
retardation ensue and modify the transition matrix elements. The effects
(B) and (C) are intimately related\cite{friar}, and early
work\cite{temp1}\cite{temp2} on relativistic corrections emphasized (C).

Relativistic Hamiltonians are not as widely known as, for example, the
relativistic field theory, though many researchers\cite{glockle,hajduk}
have utilized them. In order to study the relation between relativistic
Hamiltonians and relativistic field theory we consider interactions between
point Dirac particles coupled to scalar and vector fields. The
one-meson-exchange scattering of two particles, commonly studied with
Feynman's method, is discussed in section IV. The $\delta v({\bf P})$ is
necessary to obtain correct scattering amplitudes in frames in which
${\bf P}\neq 0$.

In the mean-field limit the problem of infinite matter consisting of
Dirac particles coupled to scalar and vector fields has been solved
by Walecka\cite{walecka}. In section V we study this problem with
relativistic Hamiltonian using the Hartree approximation corresponding to
the mean-field limit. This study demonstrates the importance of consistently
treating the relativistic effects in
$\tilde v_{ij}$, $\delta v_{ij}({\bf P}_{ij})$, $\tilde V_{ijk}$, etc.

In section VI we discuss relativistic Hamiltonians for nuclei and suggest
that they should contain the relativistic correction
for the one-pion-exchange contribution to $\tilde v_{ij}$ that has
been neglected in many realistic models of $v_{ij}$. The concluding
section VII also treats, for the sake of completeness, the motion
of the nucleus as a whole.

Many, but not all of the results given in this pedagogical article
have been published in disparate places; however, no comprehensive
discussion of them exists. In view of the recent successes in calculating
the behavior of light nuclei from realistic Hamiltonians, such a discussion
may be both useful and timely.

\section{Calculation of $\delta \lowercase{v}({\bf P})$}
Equation (1.7) for $\delta v({\bf P})$ has been obtained by Foldy\cite{foldy}
and Friar\cite{friar} using general principles of relativistic quantum
mechanics as illustrated here. Consider a system of two particles (1 and 2),
each with spin ${\bf s}$ and mass $m$. When the momentum and angular momentum
generators of the Poincar\'{e} group are chosen in the conventional fashion,
they are independent of the interaction:
\begin{eqnarray}
{\bf P}&=&{\bf p}_1+{\bf p}_2, \\
{\bf J}&=&\left({\bf r}_1\times{\bf p}_1\right)+{\bf s}_1+\left({\bf
r}_2\times{\bf p}_2\right)+{\bf s}_2,
\end{eqnarray}
while the Hamiltonian $H$ and the boost ${\bf K}$ will have interaction terms:
\begin{eqnarray}
H=H_0+H_{I}, \\
{\bf K}={\bf K}_0+{\bf K}_{I}.
\end{eqnarray}
These generators must obey the commutation relations of the
Poincar\'{e} group:
\begin{eqnarray}
\comm{P_i,P_j} & = & \comm{P_i,H} = \comm{J_i,H} = 0, \\
\comm{J_i,X_j} & = & i\epsilon_{ijk}X_k, \ \ \ \ \mbox{for} \ \ {\bf X}={\bf
J}, {\bf P}, {\bf K}, \\
\comm{K_i,K_j} & = & -i\epsilon_{ijk}J_k, \\
\comm{K_i,P_j} & = & iH\delta_{ij}, \\
\comm{K_i,H} & = & iP_i.
\end{eqnarray}
These relations are obviously satisfied by the $H_0$ and ${\bf K}_0$ of the
noninteracting system. The last two commutators in (2.5) require that
$H_{I}$ be translationally and rotationally invariant, while the
commutators in (2.6) for ${\bf X}={\bf K}$ require that ${\bf K}_{I}$ be a
spatial vector. Subtracting the contributions of the noninteracting parts
from commutators (2.7--9) gives:
\begin{eqnarray}
& & \comm{K_{I,i}, K_{0,j}} + \comm{K_{0,i}, K_{I,j}} + \comm{K_{I,i},K_{I,j}}
= 0, \\
& & \comm{K_{I,i}, P_j} = iH_{I}\delta_{ij}, \\
& & \comm{{\bf K}_0, H_{I}} = \comm{H_0, {\bf K}_{I}} + \comm{H_{I}, {\bf
K}_{I}}.
\end{eqnarray}

It is convenient to expand $H$ and ${\bf K}$ in powers of $1/m^2$. The
expansions for $H_0$ and ${\bf K}_0$ are well known\cite{foldy}:
\begin{eqnarray}
H_0 & = & 2m+\frac{1}{2m}\left(p_1^2+p_2^2\right) + \ \cdots, \\
{\bf K}_0 & = & t{\bf P}+2m{\bf R}+\frac{1}{2m}\left[\frac{1}{2}\left({\bf
r}_1p_1^2+p_1^2{\bf r}_1+{\bf r}_2p_2^2+p_2^2{\bf r}_2\right)-{\bf
s}_1\times{\bf p}_1-{\bf s}_2\times{\bf p}_2\right]+\ \cdots.
\end{eqnarray}
Note that $t{\bf P}$ and $2m{\bf R}$ are of the same order and the ellipsis
represents terms of order $1/m^3$ or higher. The leading term of $H_{I}$,
denoted by $v$, is assumed to be of order $1/m$ since in systems like nuclei
the interaction energy and the nonrelativistic kinetic energy are of similar
magnitude. Thus
\begin{equation}
H_{I}=v+\delta v+\ \cdots,
\end{equation}
where $\delta v$ is of order $v/m^2$ or $1/m^3$, and the ellipsis represents
terms of order $1/m^5$ or higher. We also assume that the leading term $v$ is
independent of ${\bf P}$. The commutator (2.11) is then minimally satisfied
by taking
\begin{equation}
{\bf K}_{I}=v{\bf R}+O\left(\frac{1}{m^3}\right) \ {\mbox{and higher terms}}.
\end{equation}

The leading terms of eq. (2.12) are of order $1/m^2$. Retaining only these
we get:
\begin{eqnarray}
2m\comm{{\bf R}, \delta{v}} & = & \frac{1}{2m}\left[\,\left(p_1^2+p_2^2\right),
v{\bf R}\,\right]  \nonumber \\
& & +\frac{1}{4m}\left[\,v, \left({\bf r}_1p_1^2+p_1^2{\bf r}_1+{\bf
r}_2p_2^2+p_2^2{\bf r}_2-\boldsigma_1\times{\bf p}_1-\boldsigma_2\times{\bf
p}_2\right)\,\right],
\end{eqnarray}
where $\boldsigma=2{\bf s}$ (i.e., the $\sigma_i$ are Pauli matrices for
spin $1/2$ particles). Evaluating the commutators one obtains the basic
equation for $\delta v$:
\begin{eqnarray}
\comm{{\bf R}, \delta{v}}&=&-\frac{i}{4m^2}v{\bf P}-\frac{1}{4m^2}\comm{{\bf
r}{\bf P}\cdot{\bf p},
v}+\frac{1}{16m^2}\comm{\left(\boldsigma_1+\boldsigma_2\right)\times{\bf P}, v}
\nonumber \\
		& &+\frac{1}{8m^2}\comm{\left(\boldsigma_1-\boldsigma_2\right)\times{\bf p},
v},
\end{eqnarray}
where ${\bf p}$ is the relative momentum. This equation can not determine
$\delta v$ uniquely. We can express
\begin{equation}
\delta v=\delta v^{\prime}+ \delta v({\bf P}),
\end{equation}
where $\delta v^{\prime}$ commutes with ${\bf R}$ (i.e., it is of order
$1/m^3$ but independent of $P$). The ${\tilde v}$ is defined as:
\begin{equation}
{\tilde v}=v+\delta v^{\prime}+ {\mbox{all higher-order terms independent of
P}},
\end{equation}
and obtained from experiment using theoretical models. Eq. (2.18) can be used
to determine
$\delta v({\bf P})$ from the ${\tilde v}$, and (1.7) provides the
simplest solution of (2.18).

It is sufficient to use the $v$ of order $1/m$ in eq. (1.7) to obtain
$\delta v({\bf P})$ up to order $1/m^3$. In some cases this $v$ is a
spin-independent function of $r$, and
\begin{equation}
\delta{v}({\bf P})=-\frac{P^2}{8m^2}v(r)+\frac{1}{8m^2}{\bf P}\cdot{\bf r}{\bf
P}\cdot\boldnabla v(r)+\frac{1}{8m^2}(\boldsigma_1-\boldsigma_2)\times{\bf
P}\cdot\boldnabla v(r).
\end{equation}
As discussed in the next section, the above three terms of
$\delta v({\bf P})$ can be attributed to the relativistic energy-momentum
relation, Lorentz contraction and Thomas precession, respectively.

The boost operator ${\bf K}_I$ can have additional terms, denoted by ${\bf w}$
in [4,6], which commute with ${\bf P}$. These make unitary transformations
of the relativistic Hamiltonian, and can be chosen for convenience. The present
choice, ${\bf w}=0$, is motivated by the desire to maintain correspondence
with classical relativistic mechanics via eq. (2.21), and is suitable to
study energies of nuclear states. When ${\bf w}\neq 0$
the $\delta v({\bf P})$ has an additional term $-i[\chi_V,H_0+v]$ where
$\chi_V$ depends upon ${\bf P}$ and ${\bf w}$. The contribution of
$\delta v({\bf P})$ to the energy eigenvalue $E_I$ up to order $1/m^3$ is
given by $\langle\Psi_I|\delta v({\bf P})|\Psi_I\rangle$, where
$|\Psi_I\rangle$ are eigenstates of $(H_0+v)$. Obviously
$[\chi_V,H_0+v]$ gives zero contribution to $E_I$ for $\chi_V$ obtained
from any ${\bf w}$. When studying reactions, however, special attention must
be paid to such terms (see Section VII and ref. 10).

\section{The $\delta \lowercase{v}({\bf P})$ in Classical Relativistic
Mechanics}
The first two terms of eq. (2.21) for $\delta v({\bf P})$ were obtained using
classical considerations in ref. [7]. In relativistic classical mechanics
two particles at rest a distance ${\bf r}_0$ apart have energy:
\begin{equation}
E_0=2m+\tilde{v}({\bf r}_0)
\end{equation}
in their rest frame by definition of $\tilde{v}$. In a frame in which
these particles are moving with momentum $P$, their energy is given by:
\begin{equation}
E_{P}=2\left(m^2+\frac{P^2}{4}\right)^{1/2}+\tilde{v}({\bf r})+\delta v({\bf
P}, {\bf r})
\end{equation}
by definition of $\delta v({\bf P})$. Here ${\bf r}$ is the distance in the
moving frame;
\begin{equation}
{\bf r}={\bf r}_0-\frac{\left({\bf P}\cdot{\bf r}_0\right){\bf P}}{2E_{P}^2}
\end{equation}
due to Lorentz contraction. Now $E_{P}$ is also given by:
\begin{equation}
E_{P}=\left(E_0^2+P^2\right)^{1/2}.
\end{equation}
{}From equations (3.2--4) we get:
\begin{eqnarray}
\delta{v}({\bf P},{\bf r})\ =\ -\frac{P^2}{8m^2}\tilde{v}({\bf
r})+\frac{1}{8m^2}{\bf P}\cdot{\bf r}{\bf P}\cdot{\bf \boldnabla}\tilde{v}({\bf
r}).
\end{eqnarray}
Its first term is due to the relativistic relation (3.4) between $E_{P}$
and $E_0$, and the second due to Lorentz contraction (3.3). These terms
are respectively denoted by $\delta v_{RE}({\bf P},{\bf r})$ and
$\delta v_{LC}({\bf P},{\bf r})$ in ref. [16]. When the interacting particles
are spinless (3.5) gives their entire $\delta v({\bf P})$.

When the interacting particles have spin the last term of (2.21) is
generated by Thomas precession\cite{jackson}. The precession of the
spin ${\bf s}_1$ in the moving frame is given by
$-\boldnabla\tilde{v}(r)\times{\bf P}/4m^2$ up to order $1/m^2$. Thus
the Thomas precession potential for particle one is:
\begin{equation}
-\frac{1}{2}\boldsigma_1\cdot\frac{\boldnabla\tilde{v}(r)\times{\bf P}}{4m^2}\
=\ \frac{1}{8m^2}\boldsigma_1\cdot{\bf P}\times\boldnabla\tilde{v}.
\end{equation}
In the moving frame both particles have the same velocity, but their
accelerations are equal and opposite. Thus the Thomas precession
potential for the second particle is
$-\boldsigma_2\cdot{\bf P}\times\boldnabla\tilde{v}/8m^2$, giving
the total:
\begin{equation}
\delta v_{TP}({\bf P},{\bf
r})=\frac{1}{8m^2}\left(\boldsigma_1-\boldsigma_2\right)\times{\bf
P}\cdot\boldnabla\tilde{v}
\end{equation}
in agreement with the last term of (2.21).

The general $\delta v({\bf P})$ given by eq. (1.7) has additional terms
containing $\comm{\left(\boldsigma_1-\boldsigma_2\right), \tilde{v}}$ and
$\comm{{\bf r},\tilde{v}}$ when $\tilde{v}$ depends upon the spins and
the relative momentum. These do not have analogues
in classical mechanics, and some of them are discussed in section VI in the
context of the one-pion-exchange interaction. They are denoted by
$\delta v_{QM}({\bf P},{\bf r})$ in ref. [16]. The contribution of
$\delta v_{TP}({\bf P},{\bf r})$ to the binding energy of $^3$H and $^4$He
has been found to be rather small, and that of
$\delta v_{QM}({\bf P},{\bf r})$ is even smaller\cite{forest}. For example,
the contributions of $\delta v_{RE}$, $\delta v_{LC}$, $\delta v_{TP}$ and
and $\delta v_{QM}$ to the energy of triton are found to be 0.23(2),
0.10(1), 0.016(2) and -0.004(2) MeV, respectively, in ref. [7] and [16].

\section{Meson-Exchange Potentials}
The one-meson-exchange scattering amplitudes, from an initial two-nucleon
state ${\bf k}_1$, ${\bf k}_2$ to final state ${\bf k}_1^{\prime}$,
${\bf k}_2^{\prime}$, depend upon the momentum transfer ${\bf q}$:
\begin{equation}
{\bf q}={\bf k}_1^{\prime}-{\bf k}_1={\bf k}_2-{\bf k}_2^{\prime},
\end{equation}
the relative momenta:
\begin{eqnarray}
{\bf p} & = & \frac{1}{2}\left({\bf k}_1-{\bf k}_2\right), \\
{\bf p}^{\prime} & = & \frac{1}{2}\left({\bf k}_1^{\prime}-{\bf
k}_2^{\prime}\right)= {\bf p}+{\bf q},
\end{eqnarray}
and the total momentum
\begin{equation}
{\bf P}={\bf k}_1+{\bf k}_2={\bf k}_1^{\prime}+{\bf k}_2^{\prime}.
\end{equation}
They can be easily calculated for Dirac particles coupled to a scalar
field $\phi$ of mass $\mu_{S}$ and
\begin{equation}
H_{int}=G_{S}{\bar \psi}\psi\phi,
\end{equation}
or a vector field $V_{\mu}$ of mass $\mu_{V}$ and
\begin{equation}
H_{int}=G_{V}{\bar \psi}{\gamma}^{\mu}\psi V_{\mu}
\end{equation}
using well known Feynman diagram rules\cite{bjorken}.

The amplitudes for scalar and vector meson exchange are denoted by
$v_{X}({\bf q}, {\bf p}, {\bf P})$, $X=S$ and $V$, respectively. They
are expressed as:
\begin{equation}
v_{X}({\bf q}, {\bf p}, {\bf P})=\tilde{v}_{X}({\bf q}, {\bf p})+\delta
v_{X}({\bf P}, {\bf q}, {\bf p})
\end{equation}
to study relations between $\tilde{v}$ and $\delta v({\bf P})$. The
$\tilde{v}_{X}$ independent of $m$ is the familiar Yukawa amplitude denoted
by $v_{X}^0$:
\begin{eqnarray}
v_{S}^0(q) & = & -\frac{G_{S}^2}{q^2+{\mu}_{S}^2}, \\
v_{V}^0(q) & = & \frac{G_{V}^2}{q^2+{\mu}_{V}^2}.
\end{eqnarray}
The $\tilde{v}_{X}$ containing all terms of order $1/m^2$ is also well known:
\begin{eqnarray}
\hspace{-0.2in}{\tilde v}_{S}({\bf q}, {\bf p})\!&\!=\!&\!v_{S}^0(q)\left[
1-\frac{({\bf p}+{\bf
p}^{\prime})^2}{4m^2}-\frac{i(\boldsigma_1+\boldsigma_2)\cdot{\bf q}\times{\bf
p}}{4m^2} \right], \\
\hspace{-0.2in}{\tilde v}_{V}({\bf q}, {\bf p})\!&\!=\!&\!v_{V}^0(q)\left[
1+\frac{({\bf p}+{\bf
p}^{\prime})^2}{4m^2}-\frac{q^2}{4m^2}-\frac{\boldsigma_1\!\times\!{\bf
q}\cdot\boldsigma_2\!\times\!{\bf
q}}{4m^2}+\frac{3i(\boldsigma_1\!+\!\boldsigma_2)\cdot{\bf q}\!\times\!{\bf
p}}{4m^2}\right],
\end{eqnarray}
and $\delta v_{X}({\bf P}, {\bf q}, {\bf p})$, up to order $1/m^2$, is given
by:
\begin{eqnarray}
\hspace{-0.2in}\delta v_{X}({\bf P}, {\bf q})\!&\!=\!&\!v_{X}^0(q)\left[
\frac{({\bf P}\cdot {\bf
q})^2}{4m^2\left(q^2+{\mu}_{X}^2\right)}-\frac{P^2}{4m^2}-\frac{i(\boldsigma_1-\boldsigma_2)\cdot{\bf q}\times{\bf P}}{8m^2} \right],
\end{eqnarray}
for both $X=S$ and $V$.

The first term of the above $\delta v_{X}$ comes from the energy:
\begin{equation}
\omega^2=\frac{({\bf P}\cdot{\bf q})^2}{4m^2},
\end{equation}
carried by the exchanged meson. Up to order $1/m^2$, the Dirac spinors are
given by:
\begin{eqnarray}
u(k)=\left(1-\frac{k^2}{8m^2}\right)\left( \begin{array}{c} {\chi} \\
\frac{\mbox{\footnotesize \boldmath$\sigma$}\cdot {\bf k}}{2m}\chi \end{array}
\right),
\end{eqnarray}
where $\chi$ are Pauli spinors. Their normalizations give a contribution
of $-v_{X}^0(q)P^2/8m^2$ to the $\delta v_{X}$. Only this contribution
is considered in the earlier work by Hajduk and Sauer\cite{hajduk}, and
it accounts for half of the $P^2/4m^2$ term in eq. (4.12). The other half
of the $P^2/4m^2$ term, and the last term in $\delta v_{X}$ have different
origins in the field theories for scalar and vector meson exchange.

The $\delta v_{X}({\bf P}, {\bf q})$ can as well be obtained from the
$v_{X}^0(q)$ with the general eq. (1.7). Since the $v_{X}^0(q)$ for scalar
and vector meson exchange (eq. 4.8, 9) is independent of spins, eq(1.7)
reduces to the simpler eq. (2.21). Using that equation for
$\delta v_{X}({\bf P}, {\bf r})$ we obtain:
\begin{eqnarray}
\delta v_{X}({\bf P}, {\bf q}) & = & \int e^{-i{\bf q}\cdot{\bf r}}\delta
v_{X}({\bf P}, {\bf r})\:d^3{\bf r} \nonumber \\
& = & \frac{1}{8m^2}\int e^{-i{\bf q}\cdot{\bf r}}\left[-P^2+{\bf P}\cdot{\bf
r}{\bf P}\cdot\boldnabla+\left(\boldsigma_1-\boldsigma_2 \right)\times{\bf
P}\cdot\boldnabla\right]\:v_{X}^0(r)\:d^3{\bf r}.
\end{eqnarray}
Integrating the second term by parts gives:
\begin{eqnarray}
\int e^{-i{\bf q}\cdot{\bf r}}\left({\bf P}\cdot{\bf r}\right)\,\left({\bf
P}\cdot\boldnabla\right)\, v_{X}^0(r)\:d^3{\bf r} & = & -P^2v_{X}^0(q)-{\bf
P}\cdot{\bf q}\:{\bf P}\cdot\boldnabla_{\!\!q}\ v_{X}^0(q) \nonumber \\
     & = &-P^2v_{X}^0(q)+2\ \frac{({\bf P}\cdot{\bf q})^2}{q^2+{\mu}_{X}^2}\
v_{X}^0(q)
\end{eqnarray}
Thus the first two terms of (4.15) together give the first two terms of
(4.12), while the last term of each is in agreement.

In this context eq. (1.7) appears to be more general. The $\tilde{v}_{X}$
depends upon the nature of the exchanged meson, but the relation (1.7)
between $\delta v_{X}$ and $\tilde{v}_{X}$ is independent of the nature of $X$.
As a matter of fact we expect eq. (1.7) to be useful to determine the
${\bf P}$-dependence of the interaction between two relativistic billiard
balls dominated by their structural overlap, rather than boson exchange.

\section{Relativistic Mean-Field Theory}
The problem of extended uniform matter consisting of Dirac particles
interacting with scalar and vector fields (eq. 4.5, 6) has been solved
by Walecka\cite{walecka} in the mean-field limit. The energy density of
this matter is given by:
\begin{equation}
{\cal E}=\frac{G_{V}^2}{2\mu_{V}^2}\ \rho^2+\frac{\mu_{S}^2}{2G_{S}^2}\
(m-m^*)^2+\frac{\gamma}{(2{\pi})^3}\int_0^{k_F}(k^2+m^{*2})^{1/2}\,d^3{\bf k},
\end{equation}
where $\rho$ is the density, $\gamma$ is the degeneracy of Dirac particles
and $k_F$ is their Fermi momentum. The effective mass $m^*$ is given by:
\begin{equation}
m^*=m-G_S\ \phi_0,
\end{equation}
where $\phi_0$ is the average value of the scalar field. Minimizing ${\cal E}$
with respect to variations in $\phi_0$ gives the transcendental
self-consistency equation:
\begin{equation}
m^*=m-\frac{G_S^2}{\mu_S^2}\
\frac{\gamma}{(2{\pi})^3}\int_0^{k_F}\frac{m^*}{(k^2+m^{*2})^{1/2}}\,d^3{\bf k}
\end{equation}
which is solved by expanding $m^*$ in powers of $k_F$. With $\gamma=4$
appropriate for nuclear matter, we obtain:
\begin{eqnarray}
m^*=m-\frac{G_S^2\ \rho}{\mu_S^2}\ \left[1-\frac{3\ k_F^2}{10\ m^2}+\frac{9\
k_F^4}{56\ m^4}-\frac{5\ k_F^6}{48\ m^6}+\frac{105\ k_F^8}{1408\
m^8}-\frac{3}{5}\frac{\rho\ k_F^2}{m^3}\frac{G_S^2}{\mu_S^2}\right. \nonumber
\\
\left. +\frac{144}{175}\frac{\rho\
k_F^4}{m^5}\frac{G_S^2}{\mu_S^2}-\frac{9}{10}\frac{\rho^2\
k_F^2}{\mu_S^4}\left(\frac{G_S^2}{\mu_S^2}\right)^2\right], \ \ \ \ \ \ \ \ \ \
\ \ \ \ \ \ \ \ \ \ \
\end{eqnarray}
up to order $k_F^{11}$, noting that
\begin{equation}
\rho=\frac{\gamma}{6\pi^2}k_F^3 .
\end{equation}
The energy per particle , given by ${\cal E}/\rho$, is obtained as a power
series in $k_F$ by substituting the expansion for $m^*$ in eq. (5.1).
\begin{eqnarray}
{\cal E}/\rho\ &=&\ m+\frac{3\ k_F^2}{10\ m}+\frac{G_V^2\
\rho}{2\mu_V^2}-\frac{G_S^2\ \rho}{2\mu_s^2} \nonumber \\
   & &+\left[-\frac{3\ k_F^4}{56\ m^3}+\frac{k_F^6}{48\ m^5}-\frac{15\
k_F^8}{1408\ m^7}+\frac{21\ k_F^{10}}{3328\ m^9}+\ \cdots\right] \nonumber \\
   & &+\ \frac{G_S^2}{\mu_S^2}\ \frac{\rho}{m}\left[\frac{3\ k_F^2}{10\
m}-\frac{36\ k_F^4}{175\ m^3}+\frac{16\ k_F^6}{105\ m^5}-\frac{64\ k_F^8}{539\
m^7}+\ \cdots\right] \nonumber \\
   & &+\left(\frac{G_S^2}{\mu_S^2}\ \frac{\rho}{m}\right)^2\left[\frac{3\
k_F^2}{10\ m}-\frac{351\ k_F^4}{700\ m^3}+\ \cdots\right] \nonumber \\
   & &+\left(\frac{G_S^2}{\mu_S^2}\ \frac{\rho}{m}\right)^3\left[\frac{3\
k_F^2}{10\ m}-\ \cdots\right],
\end{eqnarray}
where the ellipsis denotes terms of order $k_F^{12}$ or higher.

We can attempt to obtain this solution starting from a relativistic
Hamiltonian appropriate for this system, using the Hartree approximation
equivalent to the mean-field approximation. In the Hartree approximation
for uniform matter only $q=0$ diagonal interactions contribute
($k_i, k_j\rightarrow k_i,k_j$). Therefore the Hamiltonian required for
the Hartree approximation is much simpler than that containing the complete
$\tilde{v}_X$ and $\delta v_X$ given by equations (4.10--12). It is given by:
\begin{eqnarray}
H_R\ \mbox{(\ for Hartree, up to order $1/m^2$)}\ =\ \sum_i\sqrt{m^2+k_i^2}
\nonumber\ +\ \ \ \ \ \ \ \ \ \ \ \ \ \ \ \ \ \ \ \ \ \ \ \ \ \ \ \ \ \ \ \ \ \
\\
\frac{1}{\Omega}\sum_{i<j}\left\{ -\frac{G_S^2}{\mu_S^2}\left[1-\frac{({\bf
k}_i-{\bf k}_j)^2}{4m^2}\right]+\frac{G_V^2}{\mu_V^2}\left[1+\frac{({\bf
k}_i-{\bf k}_j)^2}{4m^2}\right]-\frac{({\bf k}_i+{\bf
k}_j)^2}{4m^2}\left(-\frac{G_S^2}{\mu_S^2}+\frac{G_V^2}{\mu_V^2}\right)\right\}.
\end{eqnarray}
The first two interaction terms come from $\tilde{v}_S$ and $\tilde{v}_V$
(eq. 4.10, 11) and the last from $\delta v_S({\bf P})$ and $\delta v_V({\bf
P})$
(eq. 4.12). The factor $1/\Omega$ is from normalization in a box of
volume $\Omega$.

The Hartree energy obtained with the complete $H_R$ should be identical
to that given by eq. (5.6) obtained with relativistic mean-field theory.
The first row of (5.6) is just the energy obtained in the nonrelativistic
limit; it contains the contribution of interactions independent of $m$
in the $H_R$. The second row gives the relativistic correction to the
kinetic energy of a Fermi gas, which is contained in $H_R$. All the subsequent
terms in (5.6) are relativistic corrections to the interaction energy.

There are terms of order $1/m^2$ in the $\tilde{v}_X$ and $\delta v_X(P)$ of
$H_R$. Their Hartree expectation values can be easily obtained by using
the average values:
\begin{equation}
\overline{({\bf k}_i-{\bf k}_j)^2}=\overline{({\bf k}_i+{\bf
k}_j)^2}=\frac{6}{10}k_F^2.
\end{equation}
We find that the contributions of $1/m^2$ terms of $\tilde{v}_V$ and
$\delta v_V({\bf P})$ cancel, while those of $\tilde{v}_S$ and
$\delta v_S({\bf P})$ add to give the $1/m^2$ term (first in third row)
in (5.6). The $H_R$ given by (5.7) being valid only up to order $1/m^2$
can not yield the rest of terms, of order $1/m^3$ or higher, in eq. (5.6).

\vspace{2.5in}

The first term in the fourth row, of order $1/m^3$, is known\cite{brown}
to be the Hartree contribution of the three-body force $V_{ijk}$, shown
in Fig. 1. It is obtained by eliminating the antiparticle degrees of
freedom from $H_R$. There are two terms of order $1/m^4$ in (5.6). The
second term in the third row gives the contribution of $1/m^4$ parts of
$\tilde{v}_S$ and $\delta v_S({\bf P})$, while the first term in the fifth row
is the Hartree contribution of the four-body forces. In such cases the
relativistic Hamiltonian (1.3), with only two- and three-body forces
along with their exact boost corrections can at most account for all terms
up to $k_F^{10}$. In contrast the nonrelativistic Hamiltonian (1.1)
can reproduce terms up to $k_F^3$. When $G_S^2\rho/\mu_S^2m$ is of
order unity many-body forces give significant contributions, and the
usefulness of Hamiltonians like (1.3) diminishes. At low densities the
effects of correlations between particles can be important, and these
are more easily treated using Hamiltonians.

\section{Pion-Exchange Interactions}
The one-pion-exchange interaction between two nucleons may be calculated from
the pseudovector interaction:
\begin{equation}
H_{int}=-\frac{f}{\mu_\pi}{\bar
\psi}{\gamma}^{\mu}{\gamma}_5\tau_i\psi{\partial}_{\mu}\phi_i,
\end{equation}
where $\psi$ is a Dirac field representing nucleons, and $\phi_i$ denotes
the pion field with isospin $i$. The $v_\pi$ is calculated using standard
techniques of field theory and expressed as a sum of $\tilde{v}_\pi$ and
$\delta v_\pi({\bf P})$. Keeping terms up to order $1/m^2$ we obtain:
\begin{eqnarray}
\tilde v_\pi({\bf q},{\bf
p})=-\frac{f^2}{\mu_\pi^2(q^2+\mu_\pi^2)}\boldtau_1\cdot\boldtau_2\
\boldsigma_1\cdot{\bf q}\ \boldsigma_2\cdot{\bf q}
\left(1-\frac{p^2}{m^2}\right), \ \ \ \ \ \ \ \ \ \ \ \ \ \ \ \ \ \ \ \ \ \\
\nonumber \\
\delta v_\pi({\bf P},{\bf q},{\bf p})=-\frac{f^2\
\boldtau_1\cdot\boldtau_2}{\mu_\pi^2(q^2+\mu_\pi^2)}\left\{\
\boldsigma_1\cdot{\bf q}\ \boldsigma_2\cdot{\bf q}\left(\frac{\left({\bf
P}\cdot{\bf
q}\right)^2}{4m^2\left(q^2+\mu_\pi^2\right)}-\frac{P^2}{4m^2}\right) \right. \
\ \ \ \ \ \ \ \ \ \nonumber \\
\left. -\frac{{\bf P}\cdot{\bf q}}{8m^2}\,\left[\,\boldsigma_2\cdot{\bf q}\
\boldsigma_1\cdot({\bf P}+2{\bf p})+\boldsigma_1\cdot{\bf q}\
\boldsigma_2\cdot({\bf P}-2{\bf p})\ \right]\right\}. \ \ \ \ \ \ \ \,\ \
\end{eqnarray}
This result can also be obtained assuming pseudoscalar coupling:
\begin{equation}
H_{int}=iG{\bar \psi}{\gamma}_5\tau_i\psi\phi_i,
\end{equation}
with
\begin{equation}
G^2=\frac{4m^2f^2}{\mu_\pi^2},
\end{equation}
and it can be verified that the $\delta v_\pi$ obtained by inserting
$\tilde{v}_\pi$ in eq. (1.7) is identical to that given by (6.3).

The contribution of the $\delta v_\pi({\bf P})$ term containing
$\boldsigma_1\cdot{\bf q}\,\boldsigma_2\cdot{\bf q}$, to the binding energy
of $^3$H and $^4$He is calculated in ref. [7], and those of the rest of
the terms of $\delta v_\pi$ in [16]. However the $p^2/m^2$ term in
$\tilde{v}_\pi$ has been neglected in ref. [7] and almost all other models
of $v_{NN}$. In principle it can be as important as the $P^2/4m^2$ term
of $\delta v_\pi$.

We do not as yet have a complete understanding of the nucleon-nucleon
interaction. It is generally believed that the long-range part of the
interaction is given by one-pion-exchange, and this belief is strongly
supported by the Nijmegen analysis\cite{nijmegen} of the two-nucleon
scattering data. The one-pion-exchange interaction is responsible for
the quadrupole moment of the deuteron, and it appears to give large
contributions to the nuclear binding energy\cite{vijay}. The interaction
at shorter distance probably has comparable contributions from the
internal structure of the nucleon, $N\Delta$ box diagrams for example,
and from the exchange of heavier mesons. It is convenient to separate
the $\tilde{v}_{NN}$ into the one-pion-exchange part and the rest of it:
\begin{equation}
\tilde{v}_{NN}=\tilde{v}_\pi+\tilde{v}_R.
\end{equation}
The short-range cutoff of $\tilde{v}_\pi$ and the entire $\tilde{v}_R$ are
primarily determined by fitting the observed nucleon-nucleon scattering
data.

The available models, except for Bonn models\cite{bonn}, use only the
leading term, independent of $m$, of the $\tilde{v}_\pi$ (eq. 6.2).
Thus the $\tilde{v}_R$ in these models compensates for the neglected
$p^2/m^2$ term of $\tilde{v}_{\pi}$. However, this compensation can not
be exact since the $p^2/m^2$ term in $\tilde{v}_{\pi}$ generates a
momentum-dependent tensor force. Such a force
is not yet included in other models of $\tilde{v}_{NN}$. Attempts to refit
the $NN$ scattering data with the $\tilde{v}_\pi$ correct up to order
$1/m^2$ are in progress. These will presumably provide better empirical
models of $\tilde{v}_R$, and also be useful to study relativistic effects of
order $1/m^2$.

\section{Conclusion}
Unique identification of relativistic effects is difficult in nuclear physics
due to a lack of $\it{ab\ initio}$ understanding of nuclear forces from QCD.
Several relativistic effects are inadvertently included in the non-relativistic
Hamiltonian (1.1) via the phenomenological interactions $v_{ij}$ and $V_{ijk}$
obtained by fits to observed data. However non-relativistic Hamiltonians
do not contain several known relativistic effects. The relativistic
Hamiltonians given by eq. (1.3) seem to offer a practical method to include
these effects in nuclear many-body theory.

It is technically possible to treat the kinetic energy of nucleons
relativistically. The two-body problem can be easily solved in momentum
space and realistic models of $\tilde{v}_{NN}$ can be obtained by fitting
the scattering data. Faddeev--Yakubovsky\cite{kamada} and the quantum Monte
Carlo methods\cite{wiringa,carlson1,pieper} can be used with the
relativistic kinetic energy operator $\sqrt{m_i^2+p_i^2}$. It is difficult
to expand the square root beyond the non-relativistic term. The next term,
$-p_i^4/8m_i^3$, is attractive, and a Hamiltonian unbounded from below
results when $p_i^6$ and higher terms are neglected.

In contrast it appears to be useful to expand the $\delta v({\bf P}_{ij})$ in
powers of $P_{ij}^2/4m^2$ because the total momentum of an interacting
pair of nucleons in nuclei is generally much less than $m$. The lowest order
$\delta v({\bf P}_{ij})$ is relatively simple (eq. 1.7) and seems to be
dominated by the classical terms coming from the relativistic energy-momentum
relation and Lorentz contraction.

A relativistic many-body theory of nuclei can also be developed starting
from quantum field theory. The relativistic Hamiltonians and quantum field
theory imply the same relation between $\tilde{v}_{ij}$ and
$\delta v({\bf P}_{ij})$ dictated by the invariance of the Poincar\'{e} group.
The theory based on hadron fields also provides a theoretical framework
to construct models of $\tilde{v}_{ij}$, $V_{ijk}$ and many-nucleon
interactions\cite{coon,bonn,walecka,keister}. This framework is
certainly useful, but limited by the relatively small number of fields,
such as $N$, $\Delta$, $\pi$, $\rho$, $\omega$, $\cdots$, that can be treated.
If the internal structure of nucleons strongly influences the $\tilde{v}_{ij}$
then one would need to treat consistently a large number of hadron fields.
In this case it may be advantageous to use relativistic Hamiltonians
containing semi-phenomenological models of $\tilde{v}_{ij}$ and $V_{ijk}$
having field-theoretic long-range pion-exchange parts and shorter-range
phenomenological parts.

Finally we note that in this approach it is rather simple to describe a
nucleus moving with a velocity ${\bf V}$. Its wave function is given by:
\begin{equation}
|\Psi\rangle=e^{i{\bf K}\cdot{\bf u}}|\Psi_0\rangle,
\end{equation}
where $|\Psi_0\rangle$ describes the nucleus with zero total momentum,
\begin{equation}
{\bf u}={\bf V}\tanh^{-1}(|{\bf V}|)={\bf V}\left(1+\frac{1}{3}|{\bf
V}|^2+\cdots\right),
\end{equation}
and ${\bf K}$ is given by equations (2.4), (2.14) and (2.16). Up to order
$|{\bf V}|^2$ for a two-nucleon system at time $t=0$ we obtain:
\begin{eqnarray}
{\bf K}\cdot{\bf u}&=&{\bf R}\cdot{\bf
V}\left(2m+\frac{p^2}{m}+\tilde{v}+\frac{1}{4m}P^2+\frac{2m}{3}|{\bf
V}|^2\right)-\frac{i}{2m}{\bf P}\cdot{\bf V} \nonumber \\
& &+\frac{1}{2m}\left({\bf r}\cdot{\bf V}\right)\left({\bf P}\cdot{\bf
p}\right)-\frac{1}{2m}\left[\frac{1}{2}\left({\bf s}_1+{\bf
s}_2\right)\times{\bf P}-\left({\bf s}_1-{\bf s}_2\right)\times{\bf
p}\right]\cdot{\bf V}.
\end{eqnarray}
The ${\bf P}$ and ${\bf p}$ in the above equation are operators. Using
\begin{equation}
e^{i{\bf K}\cdot{\bf u}}\
|\Psi_0\rangle=\lim_{n\rightarrow\infty}\left(e^{i{\bf K}\cdot{\bf
u}/n}\right)^n|\Psi_0\rangle,
\end{equation}
we obtain:
\begin{eqnarray}
|\Psi\rangle&=&\left[1+\frac{1}{2}\left({\bf r}\cdot{\bf V}\right)\left({\bf
V}\cdot\boldnabla\right)+\frac{1}{2m}\left({\bf s}_1-{\bf s}_2\right)\times{\bf
p}\cdot{\bf V}\right]\left(1+\frac{1}{4}|{\bf V}|^2\right)^2 \nonumber \\
& & \times\ \exp{\left[i{\bf R}\cdot{\bf
V}\left(2m+\frac{p^2}{m}+\tilde{v}+m|{\bf V}|^2\right)\right]}\ |\Psi_0\rangle.
\end{eqnarray}
Since the energy of the two-nucleon state is given by:
\begin{equation}
E=2m+\frac{p^2}{m}+\tilde{v}+m|{\bf V}|^2,
\end{equation}
we can identify ${\bf V}E$ as the value (not operator) of the total momentum.
The factor $\left(1+{\bf r}\cdot{\bf V}\ {\bf V}\cdot\boldnabla/2\right)$
in $|\Psi\rangle$ produces the Lorentz contraction of the wave function and
the $({\bf s}_1-{\bf s}_2)$ term gives spin rotations. One of the
$\left( 1+|{\bf V}|^2/4\right)$ factors compensates for the change in
normalization due to the Lorentz contraction, while the other represents
the covariant normalization $(E/E(P=0))$ of the boosted wave
function\cite{close}. Due to the choice ${\bf w}=0$ made in section II,
only kinematical changes occur in the boosted wave function. One can show
that for OPEP in the form of eq. (6.2) the ${\bf w}$ is nonvanishing (see
eq. (A21) of ref. 10).

\acknowledgements
The authors thank Dr.\ J.\ Carlson and Professor R.\ Schiavilla for many
discussions. This work was supported in part by the U.S.\ National Science
Foundation via grant PHY 89--21025. The work of J.L. Friar was performed
under the auspices of the U.S. Department of Energy.

\end{document}